\documentclass{pasj00}

\def\commenta{$^*$}
\def\commentb{$^\dagger$}
\def\commentc{$^\ddagger$}

\def\submitted{submitted}

\DeclareAbbreviation\AAHam{Astron. Abh. Hamburg. Sternw.}
\DeclareAbbreviation\AARv{Astron. Astrophys. Rev.}
\DeclareAbbreviation\AAS{American Astron. Soc. Meeting Abstracts}
\DeclareAbbreviation\AcA{Acta Astron.}
\DeclareAbbreviation\actaa{Acta Astron.}
\DeclareAbbreviation\Afz{Astrofizika}
\DeclareAbbreviation\AGAb{Astronomische Gesellschaft Abstract Ser.}
\DeclareAbbreviation\an{Astron. Nachr.}
\DeclareAbbreviation\AnAp{Annales d'Astrophysique}
\DeclareAbbreviation\AnTok{Tokyo Astron. Obs. Annals, Sec. Ser.}
\DeclareAbbreviation\Ap{Astrophysics}
\DeclareAbbreviation\ARep{Astron. Rep.}
\DeclareAbbreviation\AstBu{Astrophys. Bull.}
\DeclareAbbreviation\ATel{Astron. Telegram}
\DeclareAbbreviation\ATsir{Astron. Tsirk.}
\DeclareAbbreviation\AcApS{Acta Astrophys. Sinica}
\DeclareAbbreviation\AstL{Astron. Lett.}
\DeclareAbbreviation\BaltA{Baltic Astron.}
\DeclareAbbreviation\BANS{Bull. of the Astron. Institutes of the Netherlands Suppl. Ser.}
\DeclareAbbreviation\BASI{Bull. Astron. Soc. India}
\DeclareAbbreviation\BeSN{Be Newslett.}
\DeclareAbbreviation\BHarO{Harvard Coll. Obs. Bull.}
\DeclareAbbreviation\CBET{Cent. Bur. Electron. Telegrams}
\DeclareAbbreviation\ChJAA{Chinese J. of Astron. and Astrophys.}
\DeclareAbbreviation\caa{Chinese J. of Astron. and Astrophys.}
\DeclareAbbreviation\CoAsi{Asiago Contr.}
\DeclareAbbreviation\CoSka{Contributions of the Astronomical Observatory Skalnat\'e Pleso}
\DeclareAbbreviation\GCN{GRB Coord. Netw. Circ.}
\DeclareAbbreviation\ErgAN{Erg. Astron. Nachr.}
\DeclareAbbreviation\ibvs{IBVS}
\DeclareAbbreviation\IEEEP{IEEE Proc.}
\DeclareAbbreviation\JAD{J. Astron. Data}
\DeclareAbbreviation\JAVSO{J. American Assoc. Variable Star Obs.}
\DeclareAbbreviation\JBAA{J. Br. Astron. Assoc.}
\DeclareAbbreviation\JPhCS{J. of Physics Conference Series}
\DeclareAbbreviation\JPSJ{J. Phys. Soc. Japan}
\DeclareAbbreviation\JSARA{J. of the Southeastern Assoc. for Research in Astron.}
\DeclareAbbreviation\LowOB{Lowell Obs. Bull.}
\DeclareAbbreviation\MitAG{Mitteil. der Astronom. Gesell. Hamburg}
\DeclareAbbreviation\MitVS{Mitteil. Ver\"{a}nderl. Sterne}
\DeclareAbbreviation\MmSAI{Mem. Soc. Astron. Ital.}
\DeclareAbbreviation\Msngr{Messenger}
\DeclareAbbreviation\NewA{New Astron.}
\DeclareAbbreviation\na{New Astron.}
\DeclareAbbreviation\NewAR{New Astron. Rev.}
\DeclareAbbreviation\NInfo{Nauchnye Informatsii}
\DeclareAbbreviation\OAP{Odessa Astron. Publ.}
\DeclareAbbreviation\Obs{Observatory}
\DeclareAbbreviation\OEJV{Open Eur. J. on Variable Stars}
\DeclareAbbreviation\PASA{Publ. Astron. Soc. Australia}
\DeclareAbbreviation\PASAu{Publ. Astron. Soc. Australia}
\DeclareAbbreviation\PAZh{Pis'ma AZh}
\DeclareAbbreviation\POBeo{Publ. de l'Observatoire Astronomique de Beograd}
\DeclareAbbreviation\PCCP{Phys. Chem. Chem. Phys.}
\DeclareAbbreviation\PhR{Phys. Rep.}
\DeclareAbbreviation\PVSS{Publ. Variable Stars Sect. R. Astron. Soc. New Zealand}
\DeclareAbbreviation\PZ{Perem. Zvezdy}
\DeclareAbbreviation\PZP{Perem. Zvezdy, Prilozh.}
\DeclareAbbreviation\QJRAS{QJRAS}
\DeclareAbbreviation\RA{Ricerche Astronomiche}
\DeclareAbbreviation\RMxAA{Rev. Mexicana Astron. Astrof.}
\DeclareAbbreviation\RvMA{Reviews of Modern Astron.}
\DeclareAbbreviation\SASS{Society for Astronom. Sciences Ann. Symp.}
\DeclareAbbreviation\Sci{Science}
\DeclareAbbreviation\SPIE{SPIE Proc.}
\DeclareAbbreviation\SvA{Soviet Astronomy}
\DeclareAbbreviation\SvAL{Soviet Astronomy Letters}
\DeclareAbbreviation\VeSon{Ver\"{o}ff. Sternw. Sonneberg}
\DeclareAbbreviation\VSOLJBul{VSOLJ Variable Star Bull.}
\DeclareAbbreviation\yCat{VizieR Online Data Catalog}
\DeclareAbbreviation\ZA{Z. Astrophys.}

\def\IAUColloq#1#2{IAU Colloq. #1, #2}

\def\PublisherCambridge{Cambridge: Cambridge University Press}
\def\PublisherKluwer{Dordrecht: Kluwer Academic Publishers}

\def\PublisherSpringer{Berlin: Springer-Verlag}

\def\PublisherWorldScientific{Singapore: World Scientific}

\SetRunningHead{Y. Osaki and T. Kato}
               {Cause of the Superoutburst in SU UMa Stars}

\Received{2012/11/19}
\Accepted{2012/12/7}

\title{The Cause of the Superoutburst in SU UMa Stars is Finally Revealed
       by Kepler Light Curve of V1504 Cygni}

\author{Yoji \textsc{Osaki}}
\affil{Department of Astronomy, School of Science, University of Tokyo,
Hongo, Tokyo 113-0033}
\email{osaki@ruby.ocn.ne.jp}

\and

\author{Taichi \textsc{Kato}}
\affil{Department of Astronomy, Kyoto University,
       Sakyo-ku, Kyoto 606-8502}
\email{tkato@kusastro.kyoto-u.ac.jp}


\maketitle

\begin{document}

\begin{abstract}
We have studied the short cadence Kepler light curve of 
an SU UMa star, V1504 Cyg, which covers a period of $\sim$ 630 d. 
All superoutbursts in V1504 Cyg have turned out to be of 
precursor-main types, and the superhump first appears near 
the maximum of the precursor. The superhumps grow smoothly 
from the precursor to the main superoutburst, showing that the superoutburst 
was initiated by a tidal instability (as evidenced by the growing superhump) 
as envisioned in the thermal-tidal instability (TTI) model proposed by 
Osaki (1989, PASJ, 41, 1005). We performed a power spectral analysis of 
the light curve of V1504 Cyg. One of the outstanding features is 
the appearance of a negative superhump extending over 
around 300 d, well over a supercycle. We found 
that the appearance of the negative superhump tends to decrease the frequency 
of occurrence of normal outbursts. Two types of supercycles 
are recognized in V1504 Cyg, which are similar to those of 
the Type L and Type S supercycles in the light curve of VW Hyi, 
a prototype SU UMa star, introduced by Smak (1985, Acta Astron., 35, 357).  
It is found that the Type L supercycle is the one accompanied by 
the negative superhump, and Type S is that without the 
negative superhump. If we adopt a tilted disk as an origin of 
the negative superhump, two types of the supercycles are understood 
to be due to a difference in the outburst interval, which is in turn 
caused by a difference in mass supply from the secondary 
to different parts of the disk. The frequency of the negative superhump 
varies systematically during a supercycle in V1504 Cyg. 
This variation can be used as an indicator of the disk-radius variation, and 
we have found that the observed disk-radius variation in V1504 Cyg fits 
very well with a prediction of the TTI model. 
\end{abstract}

\section{Introduction}

The SU UMa stars are dwarf novae in short orbital periods that 
show two distinct types of outbursts: a short normal outburst with 
a duration of a few days, and a long superoutburst with a duration of 
typically two weeks [see, \citet{war95book} and \citet{hel01book} 
for dwarf novae in general and for SU UMa stars in particular]. 
In ordinary SU UMa stars, several short normal outbursts are 
sandwiched between two long superoutbursts, and a cycle from one 
superoutburst to the next is called a supercycle. Normal 
outbursts are believed to be essentially the same as those outbursts 
observed in ordinary dwarf novae with a longer orbital period, such as 
U Gem and SS Cyg stars; they are now well understood by considering 
the thermal limit-cycle instability in the accretion disk 
(see, e.g., \cite{can93DIreview}; \cite{las01DIDNXT}). 
During the superoutburst, periodic humps, called the superhumps, 
always appear with a period slightly longer than the orbital period 
by a few percent. The superhump phenomenon is now well understood 
by considering the tidal instability (\cite{whi88tidal}; \cite{hir90SHexcess}; 
\cite{lub91SHa}); superhumps are produced by a periodic tidal 
stressing of the eccentric precessing accretion disk, which is 
in turn produced by the tidal 3:1 resonance instability between 
the accretion-disk flow and the orbiting secondary star. 

\begin{figure*}
  \begin{center}
    \FigureFile(160mm,100mm){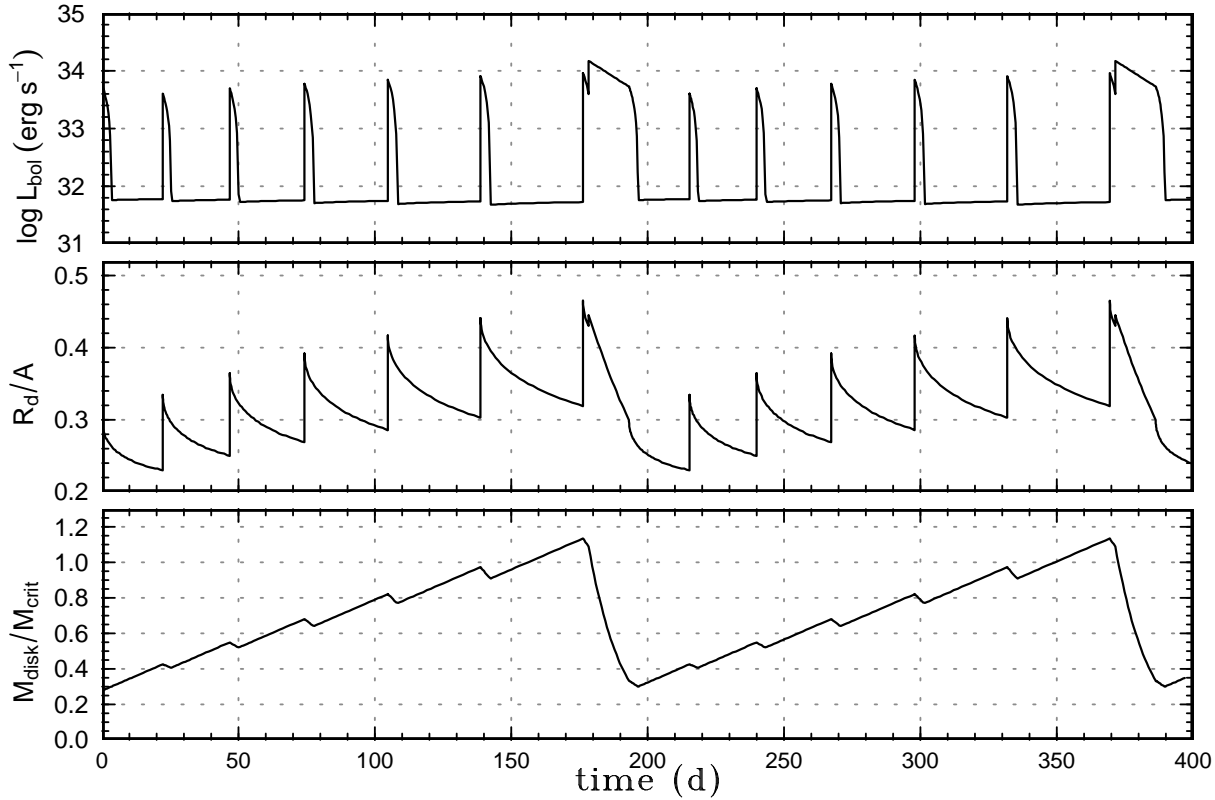}
  \end{center}
  \caption{Time evolution of an accretion disk in a supercycle based 
  on the simplified form of the TTI model, redrawn by using the result in
  \citet{osa05DImodel}.  Two supercycles are shown in the figure for clarity.
  The model parameters used are those of VW Hyi, a prototype SU UMa star.
  From the top to the bottom: bolometric light curve, 
  the disk-radius variation in units of the binary separation $A$, 
   the total disk mass, $M_{\rm disk}$, normalized by the critical mass 
  $M_{\rm crit}$ above which the disk can be tidally unstable.}
  \label{fig:osaki05fig}
\end{figure*}

As to the superoutburst and supercycle of SU UMa stars, three 
different models have so far been proposed: the thermal-tidal instability 
model advocated by Osaki, the enhanced mass-transfer model advocated by Smak, 
and the pure thermal instability model by Cannizzo. No consensus has yet been 
reached about the cause of the superoutburst. 

Besides the original planet hunting mission, NASA's Kepler observations 
\citep{Kepler} with high-accuracy photometry give an unprecedented 
opportunity to investigate variable stars. Two SU UMa stars, 
V344 Lyr and V1504 Cyg, in the Kepler field have been observed with 
the short cadence (SC) mode, and some of their light curves are 
now available to the public. In this paper, we consider the cause of 
the superoutburst in SU UMa stars using the Kepler light curve 
of one of them, V1504 Cyg, in section \ref{sec:keplerv1504cyg}. 

Before going into our study of Kepler data, we first review three models for  
superoutbursts and supercycles of SU UMa stars in 
section \ref{sec:models}.  For convenience, we summarizes in table 
\ref{tab:models} main differences among the three models and what 
we regard as being consequences of these models to be compared with
observations.

\section{Superoutburst Models}\label{sec:models} 

\subsection{Thermal-tidal Instability Model (TTI Model)}

About the superoutburst and supercycle of SU UMa stars, 
\citet{osa89suuma} proposed a model (now called the thermal-tidal 
instability model, or TTI model in short)
in which the ordinary thermal instability is coupled 
with the tidal instability [see, \citet{osa96review} for a review]. 
The TTI model is basically of 
the disk-instability variety, in which the mass-transfer rate 
from the secondary is assumed to be constant, and all variability 
is thought to be produced within the disk. 

The TTI model explains the supercycle of SU UMa stars in 
the following way. In the initial stage of the supercycle, 
the disk is assumed to be compact well below the 3:1 resonance radius. 
A successive outburst (normal outburst) causes mass accretion onto 
the central white dwarf, but the accreted mass is less than 
the mass transferred in quiescence. The mass and the angular momentum 
of the disk accumulate and the disk's outer edge therefore grows 
with a succession of normal outbursts, and the final normal outburst 
(a triggering normal outburst) brings disk's outer edge beyond 
the 3:1 resonance radius $R_{3:1}$ (where $R_{3:1}\simeq 0.46A$ 
and $A$ is the binary separation). The disk then becomes tidally unstable,  
and a circular disk is transformed to a slowly precessing 
eccentric disk. The tidal dissipation of the eccentric disk now 
enhances the mass flow in the disk, sustaining a hot state of the disk 
so as to cause a longer superoutburst. When sufficient mass is 
drained from the disk, the surface density of matter in the outer edge reaches 
the critical one, below which no hot state exists. 
The disk makes a downward transition to a cool state. The cooling front 
propagates inward, extinguishing the outburst, i.e., 
an end of the superoutburst. The eccentric disk eventually returns 
to a circular one because of the addition of matter of 
low specific angular momentum. 
Because of the enhanced tidal torque during the superoutburst stage, 
the disk becomes compact in the end. A new supercycle begins. 
This is the basic picture of the TTI model, but a minor modification 
and a further refinement of this model have been proposed by 
\citet{hel01eruma}, \citet{osa03DNoutburst},   
\citet{osa05DImodel}.

\begin{table*}
\caption{Comparison of models for superoutbursts and superhumps.}
        \label{tab:models}
\begin{center}
\begin{tabular}{p{104pt}|p{120pt}p{120pt}p{128pt}}

\hline

  \multicolumn{4}{c}{\vspace{0pt}} \\
  \multicolumn{4}{c}{{\bf Models}} \\
  \multicolumn{4}{c}{\vspace{0pt}} \\

\hline
  & thermal-tidal instability (TTI) & enhanced mass-transfer (EMT) & pure thermal instability \\
\hline
major advocator    & Osaki     & Smak    & Cannizzo \\
  \vspace{0pt} \\
mass-transfer rate from the secondary ($\dot{M}_{\rm tr}$) & constant  & variable & constant \\
  \vspace{0pt} \\
origin of superhump & eccentric disk & variable hot spot brightness & (eccentric disk?) \\
  \vspace{0pt} \\
origin of superoutburst & enhanced tidal torque & enhanced mass-transfer & wide outburst \\
  \vspace{0pt} \\
main point of the model & disk radius variation in supercycle & EMT due to irradiation of the secondary & pure thermal instability is complex enough to produce superoutburst and supercycle \\
  \vspace{0pt} \\
major premises of the model & (1) constant $\dot{M}_{\rm tr}$ & (1) enhanced mass transfer from the secondary due to irradiation heating & (1) pure thermal instability \\
         & (2) tidal instability and eccentric disk responsible for superhump and superoutburst & (2) variable and enhanced hot spot during superoutburst & (2) superhump is of secondary importance \\
         & & & (3) normal outbursts are ``inside-out'' \\
\hline

  \multicolumn{4}{c}{\vspace{0pt}} \\
  \multicolumn{4}{c}{{\bf Major consequences discussed in this paper}\commentb} \\
  \multicolumn{4}{c}{\vspace{0pt}} \\

\hline
expansion of the disk during normal outburst & Yes & Yes & \underline{No} \\ 
  \vspace{0pt} \\
outside-in type normal outburst & Yes & NA\commenta & \underline{No} \\ 
  \vspace{0pt} \\
humps during the last normal outburst before superoutburst & failed superhumps  & \underline{enhanced hot spot} & NA \\
  \vspace{0pt} \\
distinct precursor outburst & Yes & Yes & \underline{No} \\ 
  \vspace{0pt} \\
appearance of superhump  & peak of precursor 
& \underline{after peak of superoutburst}  & NA \\
  \vspace{0pt} \\
enhanced hot spot during superoutburst & No & \underline{Yes}  & No \\
\hline
  \multicolumn{4}{l}{\commentb Underlined items are not in agreement with observation or items need to be explained by the model} \\
  \multicolumn{4}{l}{(see text for details).} \\
  \multicolumn{4}{l}{\commenta Not applicable.} \\
\end{tabular}
\end{center}
\end{table*}

\citet{osa89suuma} has calculated light curves of SU UMa stars by 
using a simplified semianalytic model, consisting of 
a torus and an inviscid disk having a power-law surface-density 
distribution. Figure \ref{fig:osaki05fig} illustrates one of 
such light curves together with the disk-radius variation and 
the disk-mass variation. 
Here, we note that all outbursts in this model are of ``outside-in''
(here ``outside-in'' means an outburst in which a transition to 
the hot state first occurs at the outer part of the disk and the heating wave 
propagates inward, while ``inside-out'' does that in which a transition 
to the hot state occurs at its inner part and the heating wave propagates outward).

As far as the mass accumulated during a supercycle is concerned, 
the difference between a normal outburst just prior to a superoutburst 
and the next superoutburst is rather small, as can be seen in the lowest panel of 
figure \ref{fig:osaki05fig}. However, there is a big difference between 
these two states concerning the tidal removal of angular momentum 
from the disk. In the case of a normal outburst, the disk's outer edge 
is below the tidal 3:1 resonance radius, and thus the tidal removal of 
angular momentum is very ineffective. On the other hand, the disk's 
outer edge exceeds the tidal 3:1 resonance radius in the last 
normal outburst or the triggering outburst (which is a part of 
precursor of a superoutburst). The tidal instability then develops, 
which is observationally seen as growing superhumps.  
As discussed in \citet{osa03DNoutburst} and \citet{osa05DImodel}, 
in the case of no effective tidal removal of angular momentum from 
the disk, an expansion of the disk due to the thermal instability 
immediately returns the surface density at the outer edge of 
the disk to the critical density below which there exists no hot state. 
The disk then makes a downward transition to the cool state and 
the cooling wave propagates inward and the outburst is short. 
This is the reason why the normal outburst just prior to a superoutburst 
is so short, even if mass is sufficiently accumulated in the disk.

The key point of the TTI model lies in the disk-radius variation 
shown in the middle panel of figure \ref{fig:osaki05fig}. This must be 
tested observationally but it had been difficult before the Kepler 
observations 

\citet{ich93SHmasstransferburst} have performed light-curve 
simulations of the superoutburst and supercycle of SU UMa stars by 
using a one-dimensional numerical code based on the TTI model. 
Two different prescriptions for the viscosity in the cold state 
(i.e., $\alpha_{\rm cold}$) have been examined there.
One is a case of outside-in normal outbursts where  
the $\alpha$ viscosity has a radial dependence 
(called there ``case A'');  the other is an 
inside-out case (``case B'') where  
the $\alpha$ viscosity is constant with respect to the radius. 
It has been shown that quiescent intervals between 
normal outbursts increase monotonously with the advance of 
supercycle phase in case A (see, their figure 1) while they 
stay almost constant during the supercycle in case B 
(see, their figure 2). This point will be discussed later together with 
Kepler light curves.
\citet{bua02suumamodel} have performed light curve 
simulations by using their code, confirming that the TTI model can
account for SU UMa light curves. Two-dimensional smoothed 
particle hydrodynamics (SPH) code simulations of superhumps and 
superoutbursts of SU UMa stars were performed by \citet{mur98SH} 
and \citet{tru01DNsuperoutburst}, confirming that the superoutburst 
and superhump phenomena are a direct result of the tidal instability.
\citet{smi07SH} made 3D SPH simulations of superhumps, and found 
such enhanced tidal torques as to transfer the angular momentum from 
the disk to the binary's orbital motion when the eccentric instability 
develops, as suggested in the TTI model. 

Two alternative models to explain the superoutburst phenomenon 
have been proposed besides the TTI model . One is 
the enhanced mass-transfer model (EMT model) due to irradiation heating 
of the secondary star, advocated by Smak. The other is the 
pure thermal instability model by \citet{can10v344lyr}.

\subsection{The Enhanced Mass-Transfer Model (EMT Model)}

The enhanced mass-transfer model for superoutbursts of SU UMa stars 
was first proposed by \citet{vog83lateSH}, and discussed 
by \citet{osa85SHexcess}. In this model, the superoutburst of 
SU UMa stars is produced by enhanced mass transfer from the secondary star, 
which is in turn caused by irradiation heating of the secondary star. 
\citet{osa85SHexcess} proposed an EMT model, a model of 
the irradiation-induced mass-overflow instability as a possible 
cause of superoutbursts in SU UMa stars. However, \citet{osa96review} 
abandoned this model later in favor of the TTI model. 
Smak pursued this model instead.  

\citet{sma91suumamodel}, \citet{sma04EMT}, \citet{sma08zcha}
has been writing a series of papers in Acta Astronomica in which 
he criticizes the TTI model, and instead he advocates the enhanced 
mass-transfer (EMT) model. In the EMT model, 
the long duration of the superoutburst of SU UMa stars is thought 
to be produced by the enhanced mass transfer from the secondary star, 
which is in turn caused by irradiation heating of the atmosphere 
of the secondary star due to ultraviolet radiation from the mass-accreting 
white dwarf and the boundary layer between the accretion disk 
and the central white dwarf. In the EMT model the superhump phenomenon 
is not directly related to the mechanism of the superoutburst. 
Rather, the superhump is thought to be produced by a modulated dissipation of 
the gas stream due to modulated mass outflow, which is in turn produced by 
periodically variable irradiation of the secondary star \citep{sma09SH}.

As \citet{sma91suumamodel}, \citet{sma96superoutburst}, 
\citet{sma00DNunsolved} 
clearly stated, he was motivated to pursue 
the EMT model because of following two seeming difficulties in the TTI model, 
to which we address ourselves in this paper: 

\begin{enumerate}

\item{Observational evidence for enhanced mass transfer:}

\citet{vog83lateSH} argued in his review of VW Hyi that 
the amplitude of its orbital humps increases in maximum or declining stage of 
normal outbursts if they occur less than 40 d before the next following 
superoutburst, and interpreted that the mass-transfer rate 
from the secondary star is enhanced.
\citet{sma91suumamodel} regarded this point as one of 
the difficulties of the TTI model. 
  
\item{Sequence of events that the TTI model predicts:}

Smak raised another question concerning the TTI model, which is connected with 
the sequence of events responsible for superoutbursts and superhumps. 
He argued that in the TTI model the sequence begins with a tidal instability, 
leading to the formation of an eccentric disk, and causing 
a major enhancement of the accretion rate. Accordingly, 
the superhump should appear at the very early phase of 
a superoutburst (certainly not later than maximum). In most cases, 
however, it appears one or two days {\it after the superoutburst maximum}. 
\end{enumerate}

As regards point 1, \citet{osa03DNoutburst} already questioned 
Vogt's interpretation of ``orbital humps'', and suspected that
observed humps could be of a superhump nature. This is the very 
controversy about the nature of observed humps.
These two problems raised by Smak will be discussed while considering 
the Kepler light curve of V1504 Cyg. 

Light-curve simulations based on the EMT model were made by 
\citet{sma91suumamodel}  
and \citet{sch04vwhyimodel}. \citet{sch04vwhyimodel} performed 
light-curve simulations in both the TTI model and the EMT model, and 
they compared their results with multiwavelength light curves of 
VW Hyi, one of the best observed SU UMa stars before the Kepler observations. 
All outbursts in their models turned out to be of the inside-out 
type because of their viscosity prescription. It was found there 
that both models can generate precursor outbursts that are more 
pronounced at shorter wavelengths, which is in agreement with observations. 
Since there are adjustable free-parameters in the numerical light-curve 
simulations for both the TTI model and the EMT model, \citet{sch04vwhyimodel}
stated that it is difficult to decide which model is better for 
explaining observations. 
However, these authors have concluded that the EMT model should be  
favored over the TTI model because the EMT model is more sensitive 
to any mass-transfer rate variation, which is supposed by these authors 
to be responsible for variations seen in different supercycles in one star 
and in different SU UMa stars.

In the EMT model, the mass-transfer rate is thought to be greatly increased 
during a superoutburst. The response of the secondary's envelope to 
enhanced mass transfer has to be considered. That is, the response of 
the mass reservoir of the envelope of the secondary will affect the recurrence 
time of the supercycle in SU UMa stars.  This effect has not been taken into 
account in the simulations of \citet{sch04vwhyimodel}, and their results are 
thus incomplete in this respect. 

\begin{table*}
\caption{Superoutbursts and supercycles of V1504 Cyg.\commenta}
        \label{tab:supercycle}
\begin{center}
\begin{tabular}{cccccccccc}
\hline
(1) SC & (2) start & (3) start & (4) end & (5) SC length  & (6) SO   & (7) SC & (8) number & (9) negative & (10) orbital \\
number & of SC\commentb & of SO\commentb & of SO\commentb & excluding SO\commentc & duration\commentc & length\commentc & of NO      &   SH         & hump \\
\hline
1 &  --  & 74.5  & 88.5 &  --   & 14   & $>$88 & $>$8 & no & no \\
2 & 88.5 & 201   & 215  & 112.5 & 14   & 126.5 &  10  & no & no \\
3 & 215  & 312   & 325  &  97   & 13   & 110   &  10  & later half & partly \\
4 & 325  & 406.5 & 419  &  81.5 & 12.5 &  94   &   6  & full & no \\
5 & 419  & 516   & 530  &  97   & 14   & 111   &   5  & full & no \\
6 & 530  &  --   &  --  &  --   &  --  &   --  &  --  & early part & later part \\
\hline
  \multicolumn{10}{l}{\commenta Abbreviations in this table: supercycle (SC), superoutburst (SO), normal outburst (NO), superhump (SH).} \\
  \multicolumn{10}{l}{\commentb BJD$-$2455000.} \\
  \multicolumn{10}{l}{\commentc Unit: d.} \\
\end{tabular}
\end{center}
\end{table*}

\begin{figure*}
  \begin{center}
    \FigureFile(160mm,100mm){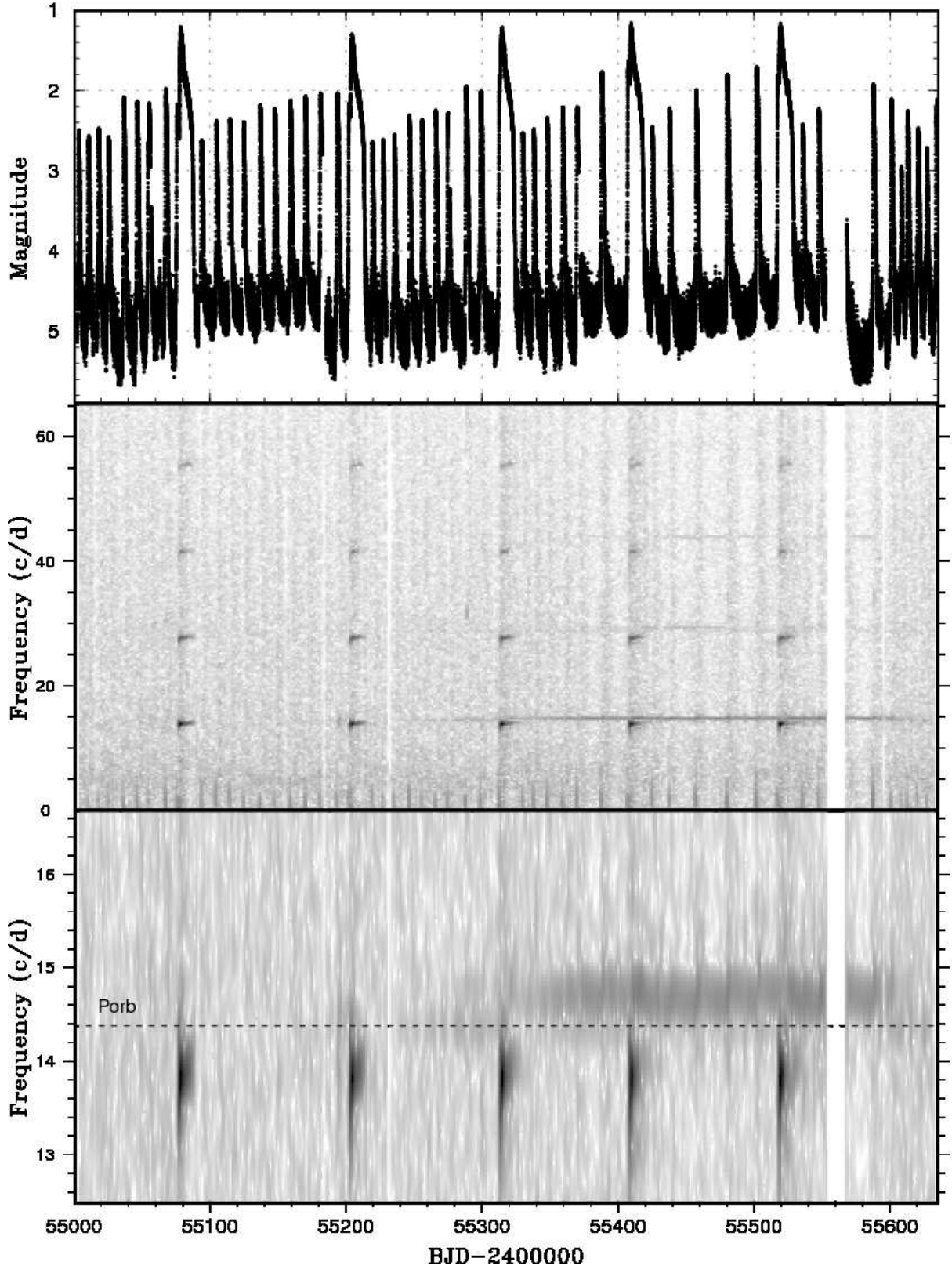}
  \end{center}
  \caption{Two-dimensional power spectrum of the Kepler light curve of 
  V1504 Cyg for the all data. From the top to the bottom, 
  (upper:) light curve; the Kepler data were binned to 0.005 d,
  (middle:) power spectrum, (lower:) its enlargement for 
  the frequency region around the orbital one. The width of 
  the moving window and the time step used are 5 d and 0.5 d,
  respectively.}
  \label{fig:v1504spec2d}
\end{figure*}

\subsection{The Pure Thermal Limit Cycle Model}
 
Let us now turn to another alternative, the pure thermal 
instability model proposed by \citet{can10v344lyr}, \citet{can12v344lyr}
and references therein. In this model, the short normal outburst and 
the long superoutburst of SU UMa stars are thought to be ``narrow''
and ``wide'' outbursts seen in SS Cyg-type dwarf novae of 
longer orbital period systems, respectively. This is an idea first proposed by 
\citet{vanpar83superoutburst}. In this model, not only the short 
normal outburst, but also the long superoutburst is explained by 
the standard thermal limit cycle instability, and the tidal instability 
is not needed to explain the long duration of the superoutburst and 
the supercycle phenomenon in SU UMa stars. The superhump is only 
an additional phenomenon, and it is of secondary importance. 
\citet{can10v344lyr} performed numerical simulations, 
demonstrating that the thermal limit cycle instability without 
any tidal instability is complex enough to produce the superoutburst and 
the supercycle of SU UMa stars in which several short outbursts are 
sandwiched by two long outbursts. 

In Cannizzo's pure thermal limit-cycle model, the short normal 
outburst is explained by an inside-out outburst in which an outburst 
is started in the inner part of the accretion disk and the heating front 
propagates outward, but it does not reach the disk's outer edge and is 
reflected in the middle of the disk as a cooling wave [see, figures 3 and 4 
of \citet{can10v344lyr}].  This type of outburst is called as ``Type Bb'' 
in Smak's classification, and it is known that the disk's outer edge does 
not expand in this type of outburst, even when an outburst occurs 
(see \cite{sma84DI}). In this model, the mass in the outer part of the disk 
is untapped during short normal outbursts, but only the mass in the inner part 
is accreted during these outbursts. The triggering outburst 
in this model is also the inside-out outburst, but this time 
the heating front traveling outward finally reaches the disk's 
outer edge, producing a long superoutburst with the viscous 
plateau stage (i.e., Smak's Type Ba outburst). 

As discussed by \citet{sma84DI}, an alternation of narrow and wide 
outbursts can be produced in the case of the inside-out outbursts. 
On the other hand, in the case of the outside-in outbursts, 
the widths of outbursts are more or less similar. 
If any normal outburst in SU UMa stars turns out to be outside-in 
based on observations, Cannizzo's model will meet a serious difficulty. 
We discuss below Cannizzo's pure thermal instability model as well.

\section{Kepler Light Curve of V1504 Cyg}\label{sec:keplerv1504cyg}

Kepler observations now open a possibility of 
discriminating between these three models of SU UMa stars from 
a purely observational point of view.
Two SU UMa-type dwarf novae, V1504 Cyg and V344 Lyr, 
were observed with the short cadence mode by the Kepler satellite. 
These light curves have already been studied by Cannizzo and his group, 
i.e., concerning quiescent superhumps \citep{sti10v344lyr}, 
numerical models of the long term light curve of V344 Lyr
\citep{can10v344lyr}, studies of both positive and negative superhumps 
in the long term light curve of V344 Lyr \citep{woo11v344lyr},  
and a study of the outburst properties of V1504 Cyg and V344 Lyr
\citep{can12v344lyr}. 
In this paper we examine the same Kepler data of V1504 Cyg, but 
from a different point of view. Data that we have used in this paper 
extend over the period from 2009 June to 2011 March. 
We discuss three problems: 
the global light curve (subsection \ref{sec:globalLC}),  
the way superoutbursts start (subsection \ref{sec:startSO}), 
and the negative superhump (subsections \ref{sec:negSH}
and \ref{sec:negpos}). 
As shown below, we have reached quite a different conclusion from that of 
\citet{can12v344lyr} concerning the nature of the superoutburst and supercycle. 

\subsection{Global Light Curves and Supercycles 
of V1504 Cyg}\label{sec:globalLC}

The Kepler light curves of V1504 Cyg and V344 Lyr, extending over 736 d 
at 1 min cadence, have been examined by \citet{can12v344lyr}; 
they have studied various correlations, such as quiescence intervals 
between normal outbursts for the period of one supercycle. Here, we examine 
the same data set of V1504 Cyg from a different standpoint. 
The data we used are those of a public release of 632 d at 1 min 
cadence. Since \citet{woo11v344lyr} examined V344 Lyr, 
we examine data of V1504 Cyg here.

The Kepler light curve of V1504 Cyg was already studied by \citet{Pdot3}. 
We summarize the basic data of V1504 Cyg obtained there; 
its orbital period is 0.069549 d (1.67 hr or 14.38 c/d),
the (ordinary or positive) superhump period $\sim$ 0.072 d (13.8 c/d), 
and the negative superhump period $\sim$ 0.068 d (14.7 c/d).

We made a two-dimensional power spectral analysis 
(the dynamic spectrum) of the light curve of V1504 Cyg, and show 
our results together with the light curve in 
figure \ref{fig:v1504spec2d}. We used the Kepler raw data
({\tt SAP\_FLUX}) during the period of 632 d 
from Barycentric Julian Date (BJD) 2455002 to 2455635.
In calculating the power spectra, we used a locally-weighted polynomial 
regression (LOWESS: \cite{LOWESS}) to Kepler magnitudes (on an arbitrary
zero-point) for removing trends resulting from 
outbursts using smoothing parameters ($f$=0.0003, $\delta$=0.2)
in R software\footnote{
   The R Foundation for Statistical Computing:\\
   $<$http://cran.r-project.org/$>$.
}.
We then estimated the pulsed flux by multiplying the residual amplitudes
and LOWESS-smoothed light curve converted to the flux scale.
In calculating the Fourier spectrum, we used a Hann window function with  
a 5 d width of the moving window, and 0.5 d as the time step.
Figure \ref{fig:v1504spec2d} shows the overall light curve in the top panel, 
power spectra in a wider frequency range in the middle, 
and their enlarged portion near the orbital frequency of the star 
in the bottom, since we are interested in this region.

Let us first look at figure \ref{fig:v1504spec2d} for light curve of 
V1504 Cyg.  The Kepler data of V1504 Cyg we use here include 
five superoutbursts and four supercycles; we summarize their main 
characteristics in table \ref{tab:supercycle}.
Here, we define the start of a supercycle as "the start of quiescence just 
after the preceding superoutburst'' and its end as "the end of the 
superoutburst.'' 
The first column (1) of table \ref{tab:supercycle} is the supercycle 
ordinal number of our data, while the next three columns give (2) BJD 
of the start of a supercycle, (3) the start of a superoutburst, 
and (4) BJD of the end of the superoutburst counted 
from BJD 2455000. Thus the date of the start of a supercycle is 
the same as that of the end of the preceding superoutburst, as can be seen 
in the table. The following three columns give the lengths in days of 
a supercycle excluding superoutburst (5), of the superoutburst duration (6), 
and of the full supercycle (7), respectively; thus the sum of the two 
columns, (5) and (6), is equal to column (7). The next column (8) 
gives number of normal outbursts during a supercycle. The last 
two columns, (9) and (10), give comments on the appearance 
(or visibility) of negative superhumps and orbital humps in 
the power spectrum of figure \ref{fig:v1504spec2d}, respectively, 
where ``no'' means no strong signals in the power spectrum. 
We have four complete supercycles No. 2--5, while supercycles No. 1 
and No. 6 are incomplete because they are lacking in the preceding 
and the following superoutbursts, respectively.

As can be seen in the light curve of figure \ref{fig:v1504spec2d} and 
in table \ref{tab:supercycle}, one of the most outstanding features in  
supercycles of V1504 Cyg is that the number of normal outbursts 
during a supercycle differs greatly from supercycle to supercycle, 
i.e., it varies from 5 (for supercycle No. 5) to 
10 (for supercycles No. 2 and No. 3) 
in the case of V1504 Cyg.  Although the duration of the supercycle differs 
among different supercycles (its average value is $\sim$ 110 d in 
V1504 Cyg), its difference is small as compared with the difference 
in the number of outbursts among different supercycles.
The duration of superoutbursts does not differ much in  
different supercycles, either. This point will be discussed later 
in this subsection.

\begin{figure*}
  \begin{center}
    \FigureFile(160mm,200mm){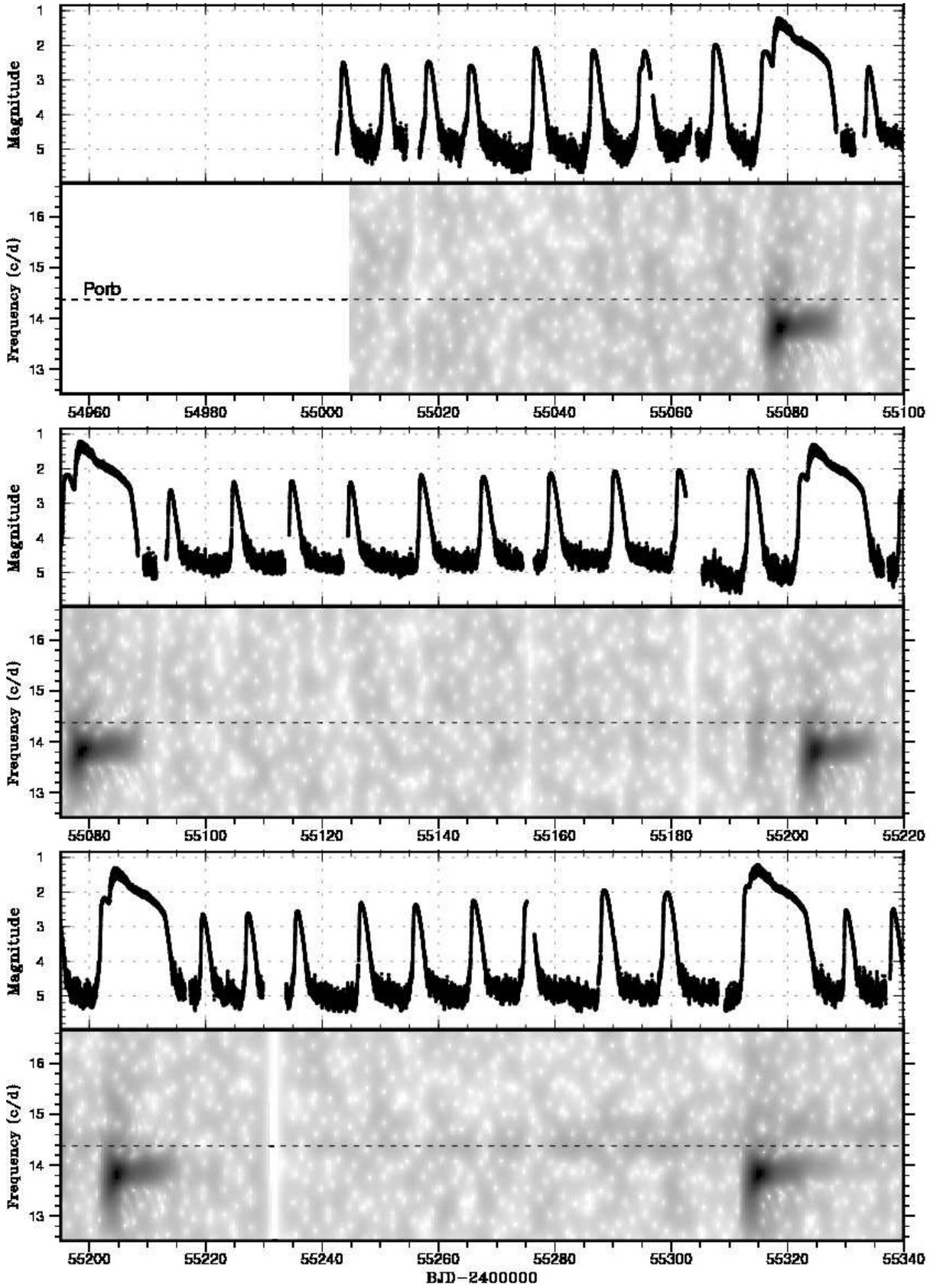}
  \end{center}
  \caption{The same as figure \ref{fig:v1504spec2d} but for six
  supercycles showing some of detailed features. The upper panel of 
  each supercycle shows the light curve and the lower panel does 
  the power spectrum around the orbital frequency region.
  The horizontal dash line represents the orbital frequency.}
  \label{fig:v1504spec2dseq}
\end{figure*}

\addtocounter{figure}{-1}
\begin{figure*}
  \begin{center}
    \FigureFile(160mm,200mm){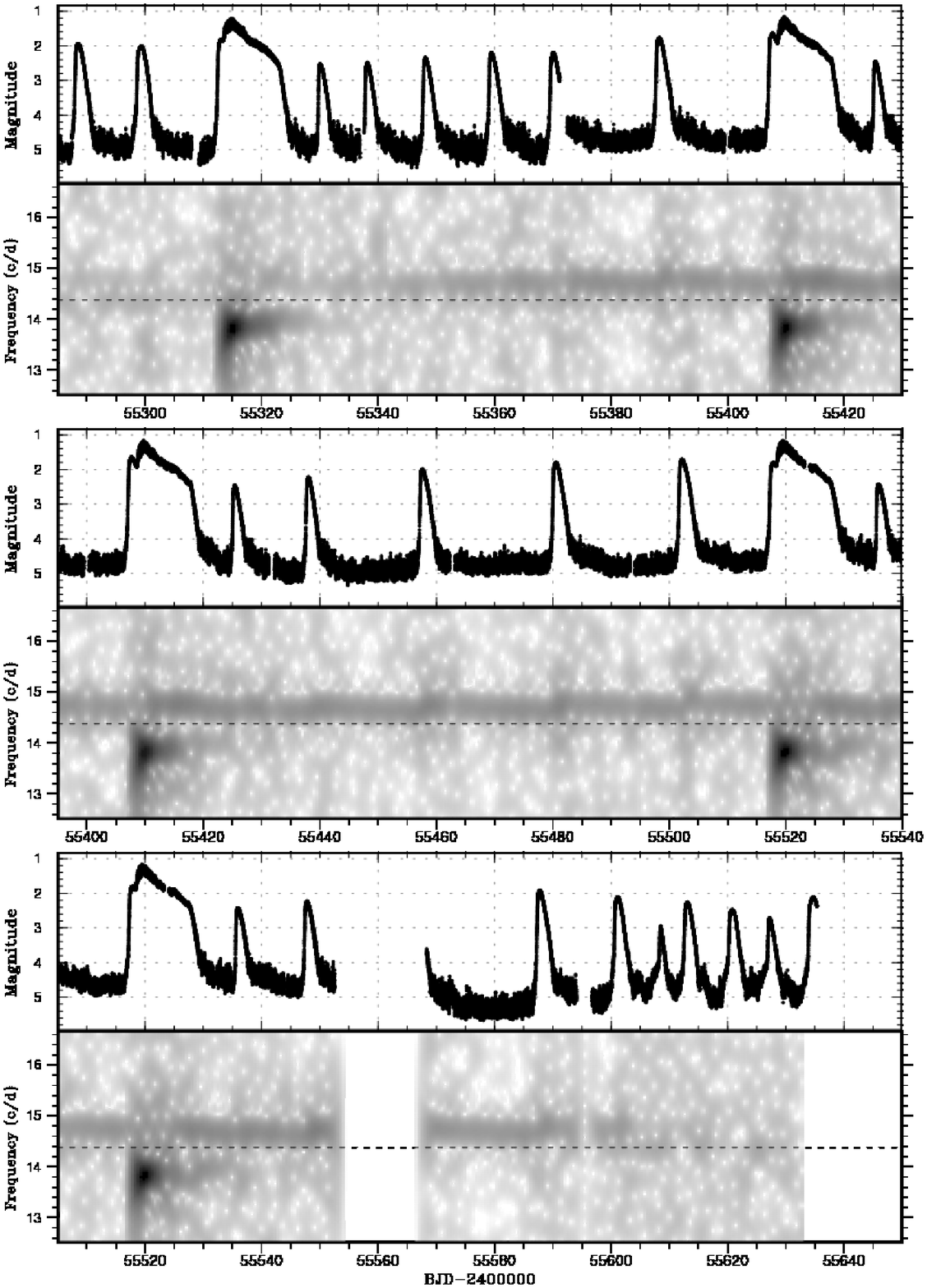}
  \end{center}
  \caption{(continued)}
\end{figure*}

Let us now examine the power spectra shown in the middle and 
lower panels of figure \ref{fig:v1504spec2d}, and also 
those of individual supercycles in figure \ref{fig:v1504spec2dseq} 
for more details where a frequency 14.38 c/d for the orbital period 
is indicated by the dashed line. We can see that the strongest signal 
in the power spectrum occurs at a frequency of around 13.8 c/d, 
corresponding to that of ordinary (positive) superhumps whenever 
a superoutburst occurs -- a well-known fact concerning SU UMa stars. 
Its higher harmonics are all visible in the middle panel of 
figure \ref{fig:v1504spec2d}, indicating the nonsinusoidal waveform 
of the positive superhump light variation. 

The most interesting aspect of 
the dynamic spectrum is the appearance of 
"negative hump'' at frequency around 14.7 c/d. 
These are humps 
with a period shorter than the binary's orbital period (a few percent shorter
than the orbital period). Although the origin of negative superhumps 
has not been yet firmly established, the standard interpretation is 
that of retrograde nodal precession of a tilted disk (e.g.,
\cite{har95v503cyg}, \cite{mon10disktilt}, \cite{woo11v344lyr}). 
Since there is no viable alternative model, we adopt the model of 
a tilted disk as a working hypothesis, and examine the Kepler data 
based on this model. In this picture the negative superhump is 
produced when the accretion disk is tilted from the binary's orbital 
plane, and its nodal line precesses retrograde, and the 
negative-superhump periodicity is produced by the synodic period between 
the retrograde precessing disk and the orbiting secondary star. 
The light variation with the negative-superhump period is thought 
to be produced by a periodic change in the dissipation of the kinetic energy 
of the gas stream from the secondary star with varying depth of potential well 
in the disk as it sweeps around the tilted disk (see \cite{woo11v344lyr}). 
When the accretion disk 
is coplanar with the orbital plane, the gas stream collides with 
the disk at the outer rim and this produces "orbital humps'' 
with the binary's orbital period.

Negative superhumps for the Kepler data of V344 Lyr have been already 
discussed from the standpoint of a tilted disk by \citet{woo11v344lyr}. 
In V344 Lyr, negative superhumps appeared both in quiescence and 
in normal outbursts; on another occasion the signature of 
the negative superhump disappeared. 
A signature of the orbital hump appeared from time to time but 
these two signals did not seem to exist simultaneously.
We can see in the lower panel of figure \ref{fig:v1504spec2d} that 
a signal of the negative superhump appeared in the middle of supercycle 
No. 3 in V1504 Cyg.  It is not clear when this signal first appeared 
but a very weak signal is recognized at around BJD 2455250. 
Its intensity increased with time,  and this signal became 
very clearly visible 
after BJD 2455350 in the later half of supercycle No. 4. 
It continued to be seen until BJD 2455600, and then tapered off.

Most interestingly, the appearance of negative
superhumps is strongly correlated with the duration 
of the quiescence interval of 
outbursts; the quiescence interval between two outbursts 
becomes longer when the negative superhump appears. That is, 
the appearance of negative superhumps tends to reduce the frequency  
of normal outbursts. The same type of phenomena 
has already been observed in other SU UMa stars with high 
mass-transfer rates, such as V503 Cyg (\cite{kat02v503cyg}, \cite{Pdot4}) 
and ER UMa \citep{ohs12eruma}.
\citet{can12v344lyr} also noticed this in the Kepler light curve of 
V344 Lyr. In our Kepler data we basically confirm the findings of 
\citet{kat02v503cyg}, \citet{Pdot4}, and \citet{ohs12eruma} that 
the existence of negative superhump suppresses frequent occurrence of 
normal outbursts.

 A similar type of difference 
in supercycles has already been noticed in light curves of VW Hyi 
(a prototype SU UMa star). \citet{sma85vwhyi} has classified 
them into two types: 
Type L supercycle in which the interval of the last two normal outbursts 
is longer than 30 d and Type S in which it is shorter than 23 d, 
where the average supercycle length of VW Hyi is $\sim$ 180 d. 
The number of normal outbursts in the Type L supercycle is small, while 
that in Type S is roughly twice as large as that of Type L. 
In Type L supercycle of VW Hyi, the quiescence interval between 
normal outbursts increases monotonously with the advance of 
the supercycle phase, while in Type S it increases till the intermediate phase 
of supercycle, but decreases in its latter half. 
\citet{sma85vwhyi} has found no correlation between these two supercycle 
types with either preceding or following superoutbursts, and with 
other supercycle properties. Thus, the origin of these two types 
remains to be a mystery. 

Here, we make use of Smak's symbols of Type L and Type S supercycles in 
our case of V1504 Cyg. We define here that the Type L 
supercycle is that in which the quiescence interval between two 
normal outbursts is relatively ``long'', but the number of 
normal outbursts during a given supercycle is small, while 
the Type S is that in which the quiescence interval between 
two normal outbursts is ``short'', but the number of normal outbursts 
during a supercycle is large. However, we note that Smak's distinction 
for Types L and S in VW Hyi concerns only the last two normal cycles, 
while it applies to most of the normal outburst intervals in our case. 
 In figure \ref{fig:v1504spec2d} we find that in V1504 Cyg, 
supercycles No. 4 and No. 5 correspond to Type L, while supercycle 
No. 2 corresponds to Type S. 

Phenomenologically speaking, we may understand from the Kepler data 
that the Type L supercycle is accompanied by 
negative superhumps while the Type S supercycle does not have 
negative superhumps. Furthermore, if we adopt a tilted disk 
as the origin of the negative superhump, we can understand how 
the two distinct types of supercycle, Type L and Type S, are produced, 
as discussed below. 

\begin{figure*}
  \begin{center}
    \FigureFile(160mm,200mm){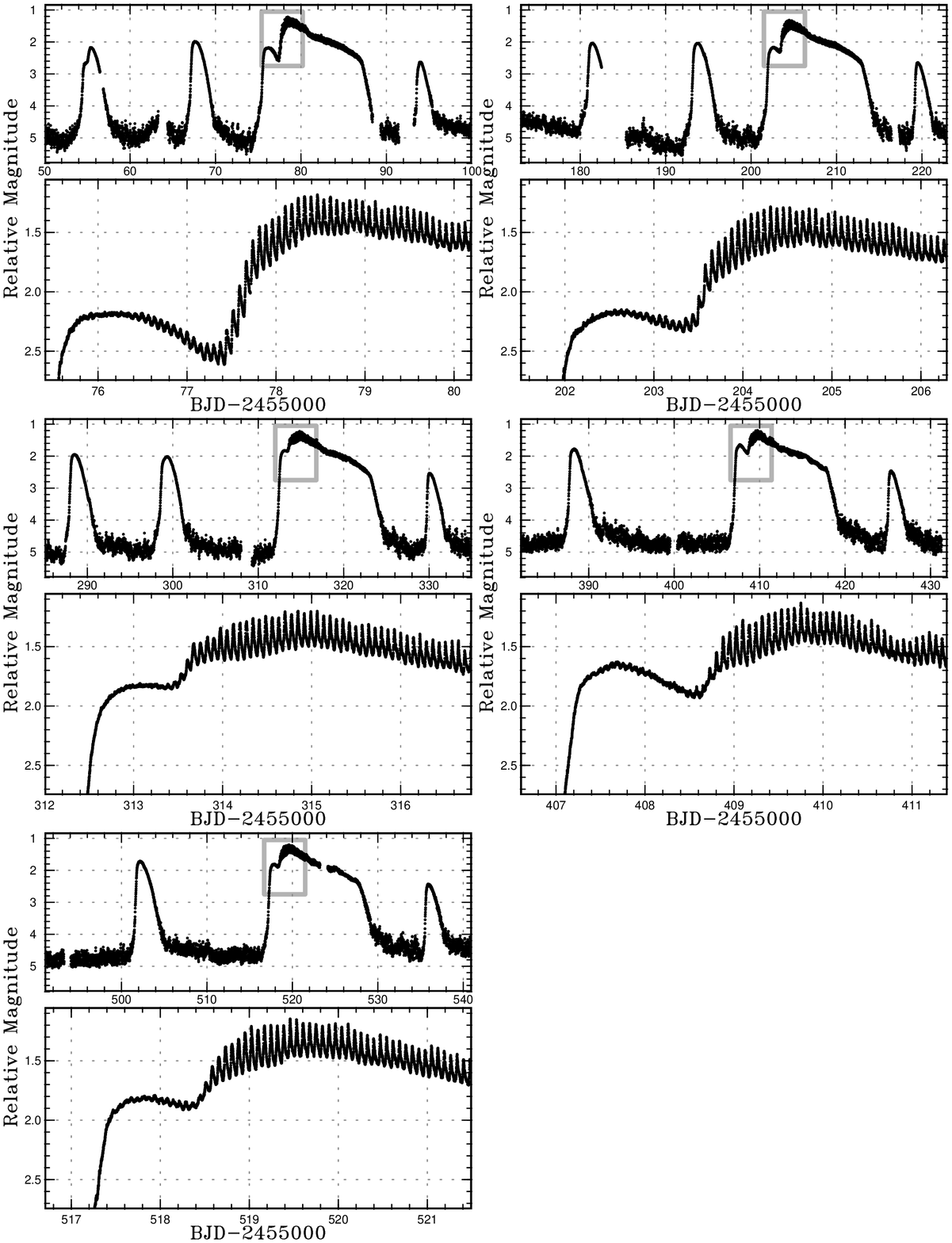}
  \end{center}
  \caption{Enlarged light curves of the five superoutbursts of V1504 Cyg 
  around their start.  The Kepler data were averaged to 0.01 d and
  0.0005 d bins in the global and enlarged figures, respectively.}
  \label{fig:v1504lclarge}
\end{figure*}

V1504 Cyg belongs to SU UMa stars with high mass-transfer rates 
since their supercycle lengths are short (i.e., its mean value of 
around 110 d), showing frequent normal outbursts. Most 
normal outbursts in V1504 Cyg are expected to be of outside-in type, 
which has been confirmed by the observed shape of the outburst light curve, 
i.e., a rapid rise to the outburst maximum and a slow decline from the maximum 
[see, \citet{sma84DI}; he calls such an outburst 
``type A outburst'']. \citet{can12v344lyr} also mentioned 
the same view. 

If the disk is co-planar with the binary orbital plane, 
the gas stream leaving the inner Lagrangian point of the secondary 
reaches the outer rim of the disk. On the other hand, if the disk 
is tilted, the gas stream sweeps over the tilted disk, but it hits 
the outer rim twice during the synodic period  
between the retrograde preccessing 
tilted disk and the orbiting secondary star (i.e., the negative-superhump 
period) and at other times it reaches the inner part of the disk.  
In such a case matter reaching 
the outer edge is effectively reduced, even if the mass-transfer rate 
from the secondary remains constant. If all normal outbursts during 
a supercycle are of outside-in types, the normal outburst 
cycle is then lengthened accordingly, because the recurrence time of 
normal outburst is essentially determined by the mass-supply rate at 
the disk's outer edge in the case of an outside-in outburst.  
This explains why the number of normal outbursts during a given supercycle 
is small in the Type L supercycle. We note here that the above discussion 
is based on an assumption of outside-in type outburst. If mass is 
supplied to the inner part of the disk via a tilted disk, 
the probability of occurrence of inside-out outburst will increase. 
Therefore we must be careful to know whether an outburst is outside-in 
or inside-out.  

Thus, it is now understood that whether the negative superhump exists or not 
makes a distinction between Type L and S supercycles in V1504 Cyg; 
 that is, the former is accompanied by negative superhumps and 
the latter is without them. However, we should note that there are 
intermediate types of supercycles 
between these two extreme cases. For instance, in our supercycle 
No. 6, in which our Kepler data are incomplete, a signal of 
negative superhumps is visible in its early phase (during a period of 
$\sim$ 70 d from BJD 2455530 to BJD 2455600) during which 
a frequent occurrence 
of normal outbursts is suppressed, but it recovered as 
the signal of negative superhumps tapered off around BJD 2455600.

Although we understand a general picture concerning 
negative superhumps and the occurrence of normal outbursts, 
a big problem remains to be solved: how on earth the 
negative superhump (i.e., a tilted disk) is produced 
and how it is maintained for a long time. We discuss 
the nature of negative superhumps in subsection 3.4 again.

\subsection{The Start of a Superoutburst}\label{sec:startSO}

Next, we examine how a superoutburst is initiated in V1504 Cyg. 
It has turned out that all superoutbursts observed in the Kepler data 
of V1504 Cyg are of precursor-main type.  Figure \ref{fig:v1504lclarge} 
shows enlarged light curves of five superoutbursts of V1504 Cyg 
in our data and one of such light curves has already been shown in 
figure 75 of \citet{Pdot3}; also see the Kepler light curves shown in 
\citet{can12v344lyr}. 
We can see very clearly in figure \ref{fig:v1504lclarge}  that 
periodic humps (which have turned out to be ``superhumps'' based on 
their period; see the power spectrum in figure \ref{fig:v1504spec2dseq}) 
appear around the maximum of the precursor (that is, 
a triggering normal outburst); they continue to grow in amplitude 
passing through the local light minimum to the main superoutburst phase, 
and the time of 
maximum amplitude of superhumps agrees fairly well with that of 
the light maximum of the superoutburst. This is exactly a picture 
envisioned in the original thermal-tidal model (TTI model) proposed 
by \citet{osa89suuma}.

Although similar phenomena have been observed before in other stars, 
e.g., V436 Cen \citep{sem80v436cen}, and QZ Vir (T Leo) \citep{kat97tleo}, 
the Kepler light curve shown here in figure \ref{fig:v1504lclarge} 
is unprecedentedly clear 
in this respect. It is quite evident from figure \ref{fig:v1504lclarge} 
that superhumps are not a result of superoutburst but rather superhumps 
(therefore the tidal instability) are most likely the cause of 
the superoutburst since superhumps start to grow near the precursor maximum.  
The superhump and superoutburst are so much entwined that one is almost 
difficult to find any interpretation other than the TTI model, that is, 
the tidal instability triggers a superoutburst in V1504 Cyg.  

Furthermore, \citet{Pdot3} have found transient low-amplitude superhumps 
in the descending branch of a normal outburst just 
prior to superoutburst No. 1 (see their figure 78). Similar superhump 
signatures are also seen in our power spectrum of 
figure \ref{fig:v1504spec2dseq} 
for normal outbursts just prior to superoutbursts No. 2 and No. 3.
This phenomenon is very well understood as the failed superhump 
(or the aborted superhump) discussed by 
\citet{osa03DNoutburst}, (see their figure 4). 

It is then natural to interpret the periodic humps in 
the case of VW Hyi discussed by \citet{vog83lateSH} as 
the same sort of phenomenon, (i.e., ``failed superhumps'').
We believe that the first question concerning about enhanced 
mass transfer prior of a superoutburst raised by 
\citet{sma96superoutburst} has been clarified by the Kepler light curve 
of V1504 Cyg.

As for the second problem raised by \citet{sma96superoutburst}, 
V1504 Cyg shows the precursor-main type superoutburst 
and the superhump appears in the precursor stage and its amplitude 
grows with the start of the main superoutburst and it reaches maximum 
almost with the superoutburst light maximum, that is, the sequence of 
events observed in V1504 Cyg is exactly as predicted by 
the original TTI model. 

We do not need to address to cases of other SU UMa stars in which 
no precursor is observed and in which superhumps appear in 
one or two days after superoutburst maximum. These superoutbursts 
are understood in the TTI model as follows: in such a superoutburst 
the disk expands to reach the tidal truncation radius, passing 
the 3:1 resonance radius during a triggering normal outburst and 
the viscous plateau stage begins first and then superhumps 
grow later, as discussed in \citet{osa03DNoutburst} and in 
\citet{osa05DImodel} where such a superoutburst is called
``Type B'' superoutburst. These superoutbursts are expected to occur 
in low mass-transfer SU UMa systems, which are more numerous in number as 
compared with high mass-transfer systems such as V1504 Cyg and V344 Lyr. 

Let us now turn our attention to the pure thermal instability model 
advocated by Cannizzo and his group. \citet{can10v344lyr} made 
numerical simulations of light curves based on the pure 
thermal-viscous limit cycle instability and compared their results 
to the Kepler light curve of V344 Lyr. The Kepler light curves of 
V344 Lyr also show precursors in superoutbursts which these authors 
called ``shoulders''. These authors succeeded in reproducing 
shoulder-like structure in their superoutburst light curve but 
the duration of the shoulder in their simulations was found to be 
too long compared with those of V344 Lyr. Our criticism to 
their pure thermal instability model for their explanation of 
the precursor is not on this point but rather on the following point. 

\begin{figure*}
  \begin{center}
    \FigureFile(160mm,100mm){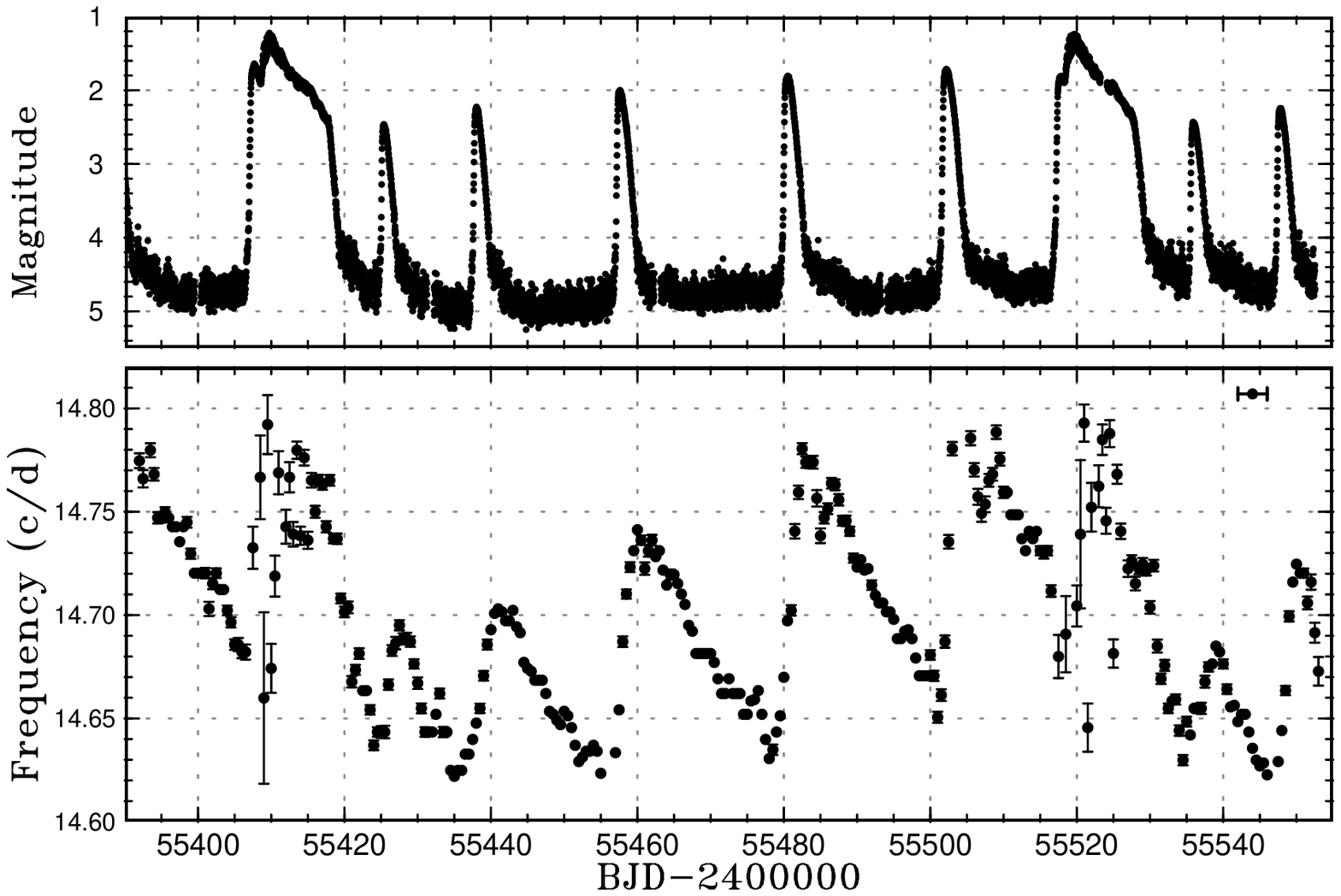}
  \end{center}
  \caption{Time evolution of frequency of the negative superhump covering 
  a complete supercycle No. 5 from BJD 2455390 to 2455550.
  The upper panel shows light curve while the lower panel does variation 
  in negative superhump frequency in units of cycle per day.
  The frequency (or period) was calculated by using the PDM method 
  with a window width of 4 d and a time step of 0.5 d. The window width 
  is indicated as a horizontal bar at the upper right corner 
  of the lower panel.
  The Kepler data were averaged to 0.0005 d bins and the error bars
  in the lower panel represent 1-$\sigma$ errors in the periods.}
  \label{fig:v1504negshvar}
\end{figure*}

In \citet{can10v344lyr} simulations, a precursor or ``shoulder'' 
is produced when the thermal instability starts in the inner part 
of the accretion disk and the heating front propagates outward 
but stagnates in the middle of disk for a little while. 
The heating front restarts to propagate outward to reach 
the outer edge of the disk which lies at the tidal truncation radius. 
The long superoutburst ensues. The shoulder is produced when 
the heating front stagnates in the middle of accretion disk 
in this model. However, the heating front which stagnates should be
never reflected back as a cooling front to produce a superoutburst
in their model. 
This is because the very outburst would have become merely 
a short normal outburst if it had been reflected. That is, 
in order to have a long superoutburst in their model, 
the heating front has to propagate all the way to the outer edge of 
disk, never being reflected. 

Kepler light curves are optical light curves but it is known that 
the dip between the precursor and main becomes deeper 
in observations of shorter wavelength, e.g., in the Voyager 
far-ultraviolet observations of VW Hyi \citep{pri87vwhyimultiwavelength}, 
and in EUV observations of OY Car \citep{mau00oycarEUVE}. 
In shorter wavelength, the precursor looks like a separate 
normal outburst. That means that the heating front has to be 
reflected as a cooling front in the precursor. This observational 
evidence clearly contradicts with the pure thermal instability model 
advocated by Cannizzo and his group, because deep dips are never 
produced in their model. 

On the other hand, in the TTI model the precursor is understood 
as a result of merging of triggering normal outburst with the main body of 
superoutburst. The degree of merging depends on individual stars and 
individual superoutbursts, sometimes a precursor is clearly separated 
from the main superoutburst and looks almost like a normal outburst, 
sometimes these two merge into one continuation. As discussed in 
\citet{osa05DImodel}, in the case of TTI model the heating front is 
reflected at the outer edge as a cooling front in the triggering 
normal outburst in the ``Type A'' superoutburst in which the outer edge of 
the disk exceeds the 3:1 resonance radius but below 
the tidal truncation radius. The tidal instability (and thus superhumps) 
ensues in this triggering normal outburst.  As superhumps grow 
in amplitude, the tidal dissipation is increased in the outer part of 
the disk and it eventually rekindles hot transition in the outer part 
of the eccentric disk, which affects most strongly to optical light.
A new separate heating front propagates from outer part inward 
(i.e., a superoutburst is initiated) while the cooling front is still 
propagating inward. Depending on the relative position of these 
two fronts, the depths of the precursor dip differ. This explains 
why the depths of the dip differ in observations of different 
wavelength and in different stars in TTI model. 

As discussed by \citet{sch04vwhyimodel}, in order to explain 
the precursor observed in SU UMa stars, the cooling front has to 
propagate inward. \citet{can10v344lyr} model cannot satisfy this 
requirement and thus the pure thermal instability model is not 
a viable model for precursor-main type superoutburst in this respect.  

\subsection{Negative Superhump and Disk Radius Variation}\label{sec:negSH}

As can be seen in the power spectrum of figure \ref{fig:v1504spec2dseq}, 
the negative superhump exists almost always during two supercycles 
No. 4 and No. 5, and some systematic variation in frequency is 
 barely visible in these figures. We thus made a detailed analysis 
of the frequency variation of the negative superhump using 
the Kepler data during the period from day 380 to day 550 covering 
a complete supercycle, No. 5. Figure \ref{fig:v1504negshvar} illustrates 
the results for the frequency variation of the negative superhump together with 
the light curve of V1504 Cyg.  We calculated its frequency using 
a data window of 4 d with a time step of 0.5 d.  We used phase-dispersion 
minimization (PDM: \cite{PDM}) for obtaining the period, and used
\citet{fer89error} and \citet{Pdot2} for estimating 1-$\sigma$ errors in
the period.  During the superoutburst plateau, we first subtracted
the signal of positive superhumps and applied the analysis to the residual
signal.  Since the amplitude of the negative superhumps was smaller
than that of positive superhumps, particularly at around the peaks of the 
superoutbursts, there were relatively large errors in estimating
the period of negative superhumps during a superoutburst.

 In figure \ref{fig:v1504negshvar} we find a characteristic variation in 
the negative-superhump frequency during a supercycle, such as is reminiscent of 
the disk-radius variation in a supercycle of the TTI model shown in 
figure \ref{fig:osaki05fig}.
In fact, if we accept 
a tilted disk model for the negative superhump, its frequency 
has turned out to be a good measure of the disk radius. 
If we use a simplified model for retrograde precession of a tilted disk, 
the frequency of the negative superhump is given by 
(see, \cite{lar98XBprecession}) 
\begin{equation}
\nu_{\rm NSH}=\nu_{\rm orb} \{ 1+(\frac{3}{7} \frac{q}{\sqrt{1+q}} 
\cos \theta )
(\frac{R_d}{A})^{3/2} \} ,
\label{equ:diskradius}
\end{equation}
where $\nu_{\rm NSH}$ and $\nu_{\rm orb}$ are the frequency for 
the negative superhump and binary orbital frequency, respectively, 
$q=M_2/M_1$ is the mass ratio of the binary, $R_d$ 
the disk radius, $A$ is the binary separation, $\theta$ is 
the tilt angle of the disk to the binary orbital plane. 
We assume $\cos \theta \simeq 1$ for a slightly tilted disk.
Furthermore if we assume the mass ratio $q=0.2$ for V1504 Cyg, we can estimate 
the disk radii, $R_d/A \simeq 0.34,$ 0.43, and 0.52 for 
$\nu_{\rm NSH}=14.6, 14.7$ and 14.8 c/d, respectively from equation
(\ref{equ:diskradius}). 

\begin{figure}
  \begin{center}
    \FigureFile(88mm,110mm){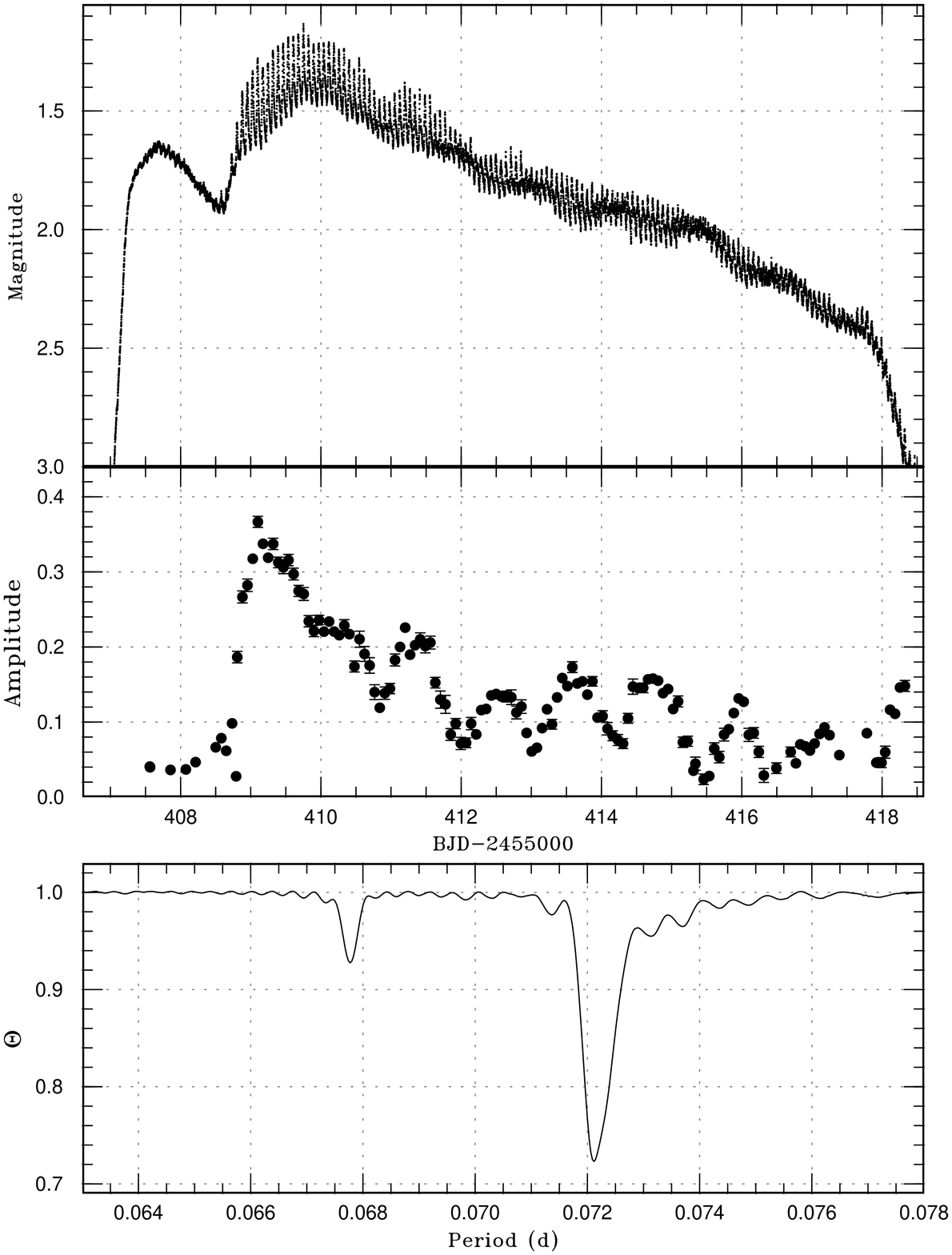}
  \end{center}
  \caption{(Upper:) A light curve of the superoutburst No. 4 showing beating 
  phenomenon between the positive superhump wave and the negative 
  superhump one.  The Kepler data were averaged to 0.0005 d bins.
  (Middle:) Amplitudes of positive superhumps.  The amplitudes were
  determined with the method in \citet{Pdot}.
  (Lower:) PDM analysis.  Both positive and negative superhumps were
  present.}
  \label{fig:v1504beat}
\end{figure}

We can see in figure \ref{fig:v1504negshvar} that the 
disk radius, $R_d$, shows a saw-tooth pattern with 
an expansion of the disk in each 
normal outburst accompanied by contraction during quiescence; 
also the average radius of the saw-tooth pattern increases with 
the advance of supercycle phase. Finally, the last normal outburst 
(which corresponds to a triggering outburst, i.e., a precursor stage of 
superoutburst)  brings the disk to a critical radius, 
(i.e., the tidal 3:1 resonance instability radius $R_{3:1}/A \sim 0.47$). 
A superoutburst 
ensues and a large amount of mass is drained from the disk during 
a superoutburst and, after the end of the superoutburst, 
the disk radius returns to a small value of radius, a picture 
exactly predicted by the TTI model as shown in 
figure \ref{fig:osaki05fig}.  Figure \ref{fig:v1504negshvar} exhibits 
clearly how the angular momentum 
is accumulated in the disk during a supercycle. 
The most important prediction of the TTI model concerns  
the disk-radius variation during a supercycle, but its observational
test was difficult until Kepler 
observations. The Kepler data of V1504 Cyg with the negative superhump 
now opens a new way to the test of this prediction. 

\subsection{Coexistence of Positive and Negative Superhumps}\label{sec:negpos}

\begin{figure}
  \begin{center}
    \FigureFile(88mm,110mm){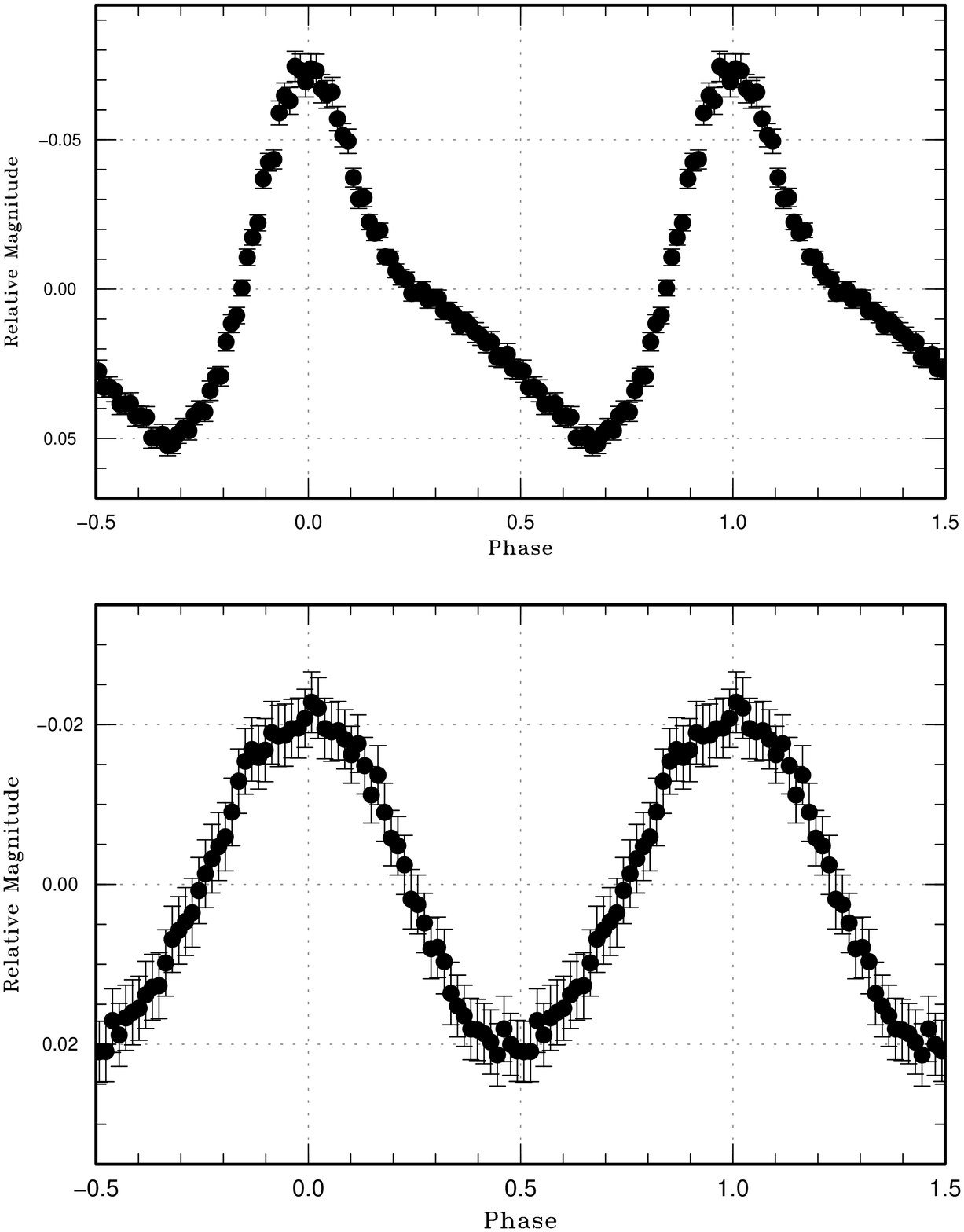}
  \end{center}
  \caption{Phase-averaged light curves of the positive superhump (upper), 
  and of the negative superhump (lower) for the superoutburst No. 4. 
  The periods used for the folding are 0.072183 d 
  and 0.067764 d, respectively.}
  \label{fig:v1504prof}
\end{figure}

\begin{figure}
  \begin{center}
    \FigureFile(88mm,70mm){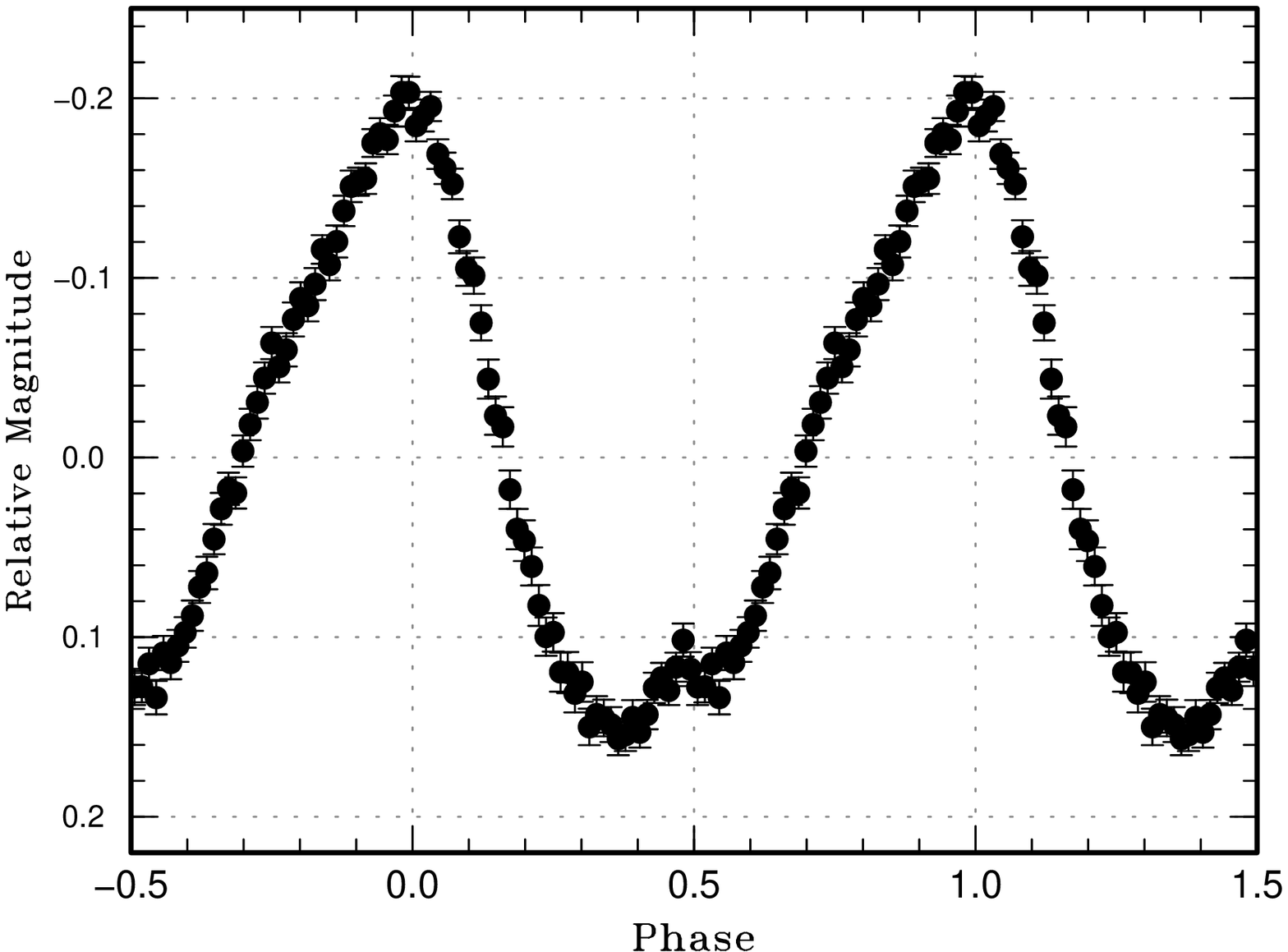}
  \end{center}
  \caption{Phase-averaged light curve of 
  the negative superhump with a period 0.068076 d during quiescence 
  for 8 d of BJD 2455440--448 in V1504 Cyg.}
  \label{fig:v1504negqui}
\end{figure}

As seen in the two-dimensional power spectra of
figure \ref{fig:v1504spec2d} and figure \ref{fig:v1504spec2dseq}, 
the positive and negative superhumps can co-exist during 
a superoutburst, such as the superoutburst No. 4 and No. 5.
Figure \ref{fig:v1504beat} illustrates a light curve of  
the superoutburst No. 4 which clearly shows a beating phenomenon of 
the positive and negative superhumps. The lowest panel of 
figure \ref{fig:v1504beat} shows a PDM diagram in this superoutburst 
in which two signals at periods 0.067764(10) d (the negative superhump) 
and 0.072183(4) d (the positive superhump) are visible. The beat period 
of these two waves is estimated to be about 1.1 d and the amplitude 
variations with this beat period are seen in the middle panel of 
figure \ref{fig:v1504beat}.

This means that the disk in 
V1504 Cyg develops tilted and eccentric form simultaneously. 
Figure \ref{fig:v1504prof} shows phase-averaged light curves of 
the positive superhumps (upper panel) and of the negative superhumps 
(lower panel) for this superoutburst. 

The light curve of the positive superhump shown in 
figure \ref{fig:v1504prof} is a typical one for the ordinary superhump 
seen during a superoutburst, that is, a rapid rise to maximum and 
slow decline sometimes accompanied with a secondary maximum. 
On the other hand, 
the light curve of the negative superhump shows more or less a sinusoidal 
waveform. For comparison, we show a phase-averaged light curve of 
the negative superhump in quiescence with a period 0.068076 d during 8 d 
of BJD 2455440--448 in figure \ref{fig:v1504negqui}.
\citet{woo11v344lyr} discussed 
the Kepler light curve of the negative superhump in quiescence for 
V344 Lyr, showing that it is approximately saw-toothed with a rise time 
roughly twice the fall time. Figure \ref{fig:v1504negqui} shows 
the same wave pattern as that of V344 Lyr. As discussed by 
\citet{woo11v344lyr} and mentioned already in subsection 
\ref{sec:globalLC}, the quiescence light curve of the negative superhump 
is understood in terms of gas stream hitting the different part of 
the disk. We discuss on a possible origin of sinusoidal waveform of the 
negative superhump during the superoutburst in Appendix.

Amplitudes (in flux) of negative superhumps in the superoutburst No. 4 and 
in neighboring quiescence are not so much different as seen from 
figures \ref{fig:v1504spec2d} and \ref{fig:v1504spec2dseq}. 
In fact, the full amplitudes are 0.04 mag and 0.35 mag,
respectively, as seen in figure \ref{fig:v1504prof} and 
figure \ref{fig:v1504negqui}. 
Since the magnitude difference between the superoutburst and quiescence 
is about 3 mag in V1504 Cyg, they are similar in flux units. This suggests 
that mass-transfer rate is not particularly enhanced during the superoutburst. 
This supports the TTI model rather than the EMT model.

\section{Summary}

(1) The long outburst of almost all of dwarf novae below the period gap 
is accompanied by superhumps.  No dwarf novae showing a superoutburst 
and a well-defined supercycle and without superhumps, 
have yet been discovered. Since the superoutburst 
and superhumps are very much entwined as demonstrated in 
figure \ref{fig:v1504lclarge}, 
no models without taking into account the tidal instability properly 
seem to be correct and we believe that the TTI model is only 
a viable model for the superoutburst and superhumps of SU UMa stars.

(2) The periodic hump observed in the decline phase of a normal outburst 
just prior to the next superoutburst is found to be of superhump property 
and the orbital hump is not particularly enhanced in this phase in V1504 Cyg. 
We conclude that no enhancement of mass transfer from the secondary star 
exists in this phase and at the start of the superoutburst. 

(3) The quiescence intervals between normal outbursts strongly depend on 
whether negative superhumps exist or not. We confirm that the 
conclusion of \citet{ohs12eruma} for ER UMa that the existence of 
negative superhumps tends to suppress the frequent occurrence of 
normal outbursts. Two types of supercycles are recognized in V1504 Cyg 
which are very similar to the Type L and Type S supercycles introduced 
by \citet{sma85vwhyi} in the case of VW Hyi. The Type L supercycle is 
a supercycle in which the number of normal outbursts is small, 
typically 5 to 6, while in the Type S supercycle it is twice as large 
(typically around 10) as that of Type L in V1504 Cyg. 
The Type L supercycle is accompanied by the negative superhump, 
while Type S is without the negative superhump.   

(4) Most of normal outbursts observed in V1504 Cyg are of the outside-in 
type. If so, lengthening the quiescence intervals when the negative 
superhump appears is understood as to be due to a decrease in mass 
supply {\it at the outer edge}, of the tilted disk as gas stream 
flows over the edge and reaches its inner part in such a case. 

(5) The frequency of the negative superhump varies systematically 
during a supercycle. If we adopt a tilted-disk model for the origin of 
the negative superhump, its frequency represents retrograde 
preccession rate of the tilted disk.  Using this variation as 
an indicator of the disk-radius variation, we found that 
the observed disk-radius variation in V1504 Cyg fits very well with 
a prediction of the TTI model.

(6) The positive and negative superhumps can coexist, seen   
as a beat phenomenon of these two waves in the light curve of 
superoutburst No. 4 in V1504 Cyg. This means that the disk can 
take eccentric and tilted form simultaneously. The amplitude of the negative 
superhump during the superoutburst is not particularly enhanced in flux units 
as compared with that of neighboring quiescence. This suggests   
no enhancement of mass-transfer rate during the superoutburst, 
which supports the TTI model rather 
than the EMT model for the origin of the superoutburst. 

(7) We summarize major consequences of the three models 
discussed in this paper in the lower part of table \ref{tab:models}. 

\medskip
 
We are grateful to Dr. Makoto Uemura for his help 
in retrieving the Kepler data. One of authors (Y. Osaki) thanks 
the organizers of a conference held in Warsaw, 2012 September
``Accretion Flow Instabilities: -- 30 years of thermal-viscous 
instability'' for encouraging him to participate, of which experience 
stimulated him to come back to research in astronomy after a 7 yr absence. 
He also would like to 
express his sincere thanks to Dr. Friedrich Meyer at 
Max-Planck-Institut f\"ur Astrophysik for encouraging him to write 
this paper and Professor Hiromoto Shibahashi for enlightening 
discussions about power spectra. 
This work was partly supported (TK) by the Grant-in-Aid for the Global 
COE Program ``The Next Generation of Physics, Spun from Universality 
and Emergence" from the Ministry of Education, Culture, Sports, Science 
and Technology (MEXT) of Japan.
We thank the Kepler Mission team and the data calibration engineers for
making Kepler data available to the public.

\quad

Note added in proof (April 30, 2013): 
We have replaced our figure 5 of the original version to a new one 
because we made a mistake of 2 d in time axis of our lower panel of 
figure 5 in the first version of a preprint, which was submitted 
to astro-ph on December 7, 2012, (arXiv:1212.1516v1), and the mistake was 
corrected in version 2 submitted on January 6, 2013  (arXiv:1212.1516v2). 
Figure 5 is now the corrected one. This mistake crept in, when we 
inadvertently used time of local frequency at that of the starting date 
of the window instead of the correct one (i.e., the middle of the window). 
We thank Dr. Smak who first noticed this error (Acta Astron, 63, 109, 2013). 
We notified him about our mistake on this point by e-mail on January 7, 2013. 

\appendix

\section{Waveform of negative superhump during a superoutburst}

The different form of the light curve of the negative superhump 
during a superoutburst from that of quiescence may suggest that 
the origin of its light variation is different between superoutburst 
and quiescence. Let us now discuss the origin of light variation of 
the negative superhump during a superoutburst. During superoutbursts 
the disk component will play an important role besides that of gas stream in 
light variation of negative superhump. 
Then a question arises why its waveform is sinusoidal if the disk component 
contributes. 

Here we propose a following explanation. Let us consider an eigenmode 
of oscillation of an accretion disk in which the unperturbed state is 
co-planar with the binary orbital plane. Here we use the cylindrical 
coordinates with $(r, \varphi, z, t)$ in the inertial frame of reference 
where the center of the coordinates is chosen to be the center of 
the disk, i.e., the central white dwarf, the azimuthal angle $\varphi$ 
is measured in the direction of the binary orbital motion, and 
the $z$-axis is that of the binary orbital plane. An eigenmode 
displacement vector is then written as 

\begin{equation}
\xi(r,\varphi, z, t)=(\xi_h,\xi_z)\cos(m\varphi-\omega t) ,
\label{equ:tiltmode}
\end{equation}
where $\xi$ is a displacement vector, 
$\omega$ is an eigenfrequency, $m$ is azimuthal wave number, 
and $\xi_h$, $\xi_z$ are horizontal and vertical displacements and 
this mode is denoted as a mode $(m, \omega)$ with azimuthal wave number 
$m$ and eigenfrequency $\omega$.

We do not discuss on the excitation mechanism of tilted disk but 
we assume here that the disk is tilted with a finite tilt angle 
$\theta$ where $\theta \ll 1$, for simplicity. 
A tilt of disk is understood 
to be one of eigenmodes with $m=1$ and $\mid\xi_h \mid \ll \mid \xi_z\mid $, 
$\xi_z\simeq \theta r$ and an eigenfrequency given by angular frequency of 
retrograde precession $\omega=\Omega_{\rm pr}<0$.
This mode itself does not produce any light variation. In order to 
understand light variation of tilted disk, we need to consider 
an interaction of this mode with the tidal field of the secondary star. 
The tidal perturbing potential in the accretion disk by the secondary star 
is expressed by 

\begin{equation}
\phi(r,\varphi, z, t) =\phi_m(r, z) \cos (m(\varphi-\Omega_{\rm orb} t)) ,
\label{equ:tidalgr}
\end{equation}
where $\Omega_{\rm orb}$ is the angular frequency of the binary orbital motion and $m=1, 2, 3, \cdots$ is 
azimuthal wave number.
The tidal perturbation potential produces deformation in the accretion disk 
which is described by (see, \cite{lub91SHa}, \cite{lub92SH}, 
\cite{kat12SHwaveresonant}) 

\begin{equation}
\xi_{\rm D}(r,\varphi, z, t)=\xi_{\rm D,m} (r, z) \cos (m(\varphi-\Omega_{\rm orb} t)) ,
\label{equ:deform}
\end{equation}
where suffix D denotes deformation.  

The wave-wave interaction of the tilt mode with the tidal deformation mode 
produces $(m-1, m\Omega_{\rm orb}-\Omega_{\rm pr})$ mode. 
In this interaction, the most important one is that of $m=1$, that is, 
a mode with $(0,\Omega_{\rm orb}-\Omega_{\rm pr})$. This mode is 
independent of the azimuthal angle $\varphi$ and the time dependence of 
$\sin \{(\Omega_{\rm orb}-\Omega_{\rm pr})t+\delta_0\}$, 
where $\delta_0$ is a constant phase. That is, this mode produces 
a sinusoidal time variation with the negative superhump period 
$P_{\rm NSH}=2\pi/(\Omega_{\rm orb}-\Omega_{\rm pr})$ 
where $\Omega_{\rm pr}<0$ is retrograde nodal 
precession rate of a tilted disk. 
This can produce light variation even with 
the pole-on geometry, i.e., in the case of inclination angle $i=0$. 
Light variation with the negative superhump period observed during 
a superoutburst in V1504 Cyg can be explained by
this disk component produced by 
wave-wave interaction between a tilt mode and the tidal deformation 
with $m=1$ besides the gas stream component.

\end{document}